\documentclass[aps,twocolumn]{revtex4}
\usepackage{graphicx}
\usepackage{latexsym}

\newcommand{\be}{\begin{equation}}
\newcommand{\ee}{\end{equation}}
\newcommand{\bea}{\begin{eqnarray}}
\newcommand{\eea}{\end{eqnarray}}

\begin{document}
\title{On the Limits of Analogy Between Self-Avoidance and Topology-Driven Swelling of Polymer Loops}

\author{N.T. Moore, A.Y. Grosberg}
\affiliation{Department of Physics, University of Minnesota, Minneapolis, MN 55455, USA}
\date{\today}

\begin{abstract}
The work addresses the analogy between trivial knotting and excluded volume in looped polymer
chains  of moderate length, $N<N_0$, where the effects of knotting are small.
A simple expression for the swelling seen in trivially knotted loops is described and
shown to agree with simulation data.  Contrast between this expression and the
well known expression for excluded volume polymers leads to a graphical mapping of
excluded volume to trivial knots, which may be useful for understanding where the analogy
between the two physical forms is valid.
The work also includes description of a new method for the computational generation of
polymer loops via conditional probability.  Although computationally intensive, this
method generates loops without statistical bias, and thus is preferable to other loop generation
routines in the region $N<N_0$.
\end{abstract}

\maketitle

\section{Introduction: Formulation of the Problem}

The last few years have seen significant work addressing the
effects of knotting on looped polymer chains. Of interest to
mathematicians and physicists for good part of nineteenth and most
of the twentieth centuries, knots were first seen by W. Thomson as
a way to understand the nature of atoms \cite{knotted_vorticies},
and more recently as the basis for string theory.  On the
biological front, knots have been observed in,
\cite{JBiolChem_1985,probability_DNA_knotting}, and tied into,
\cite{tie_knot_into_DNA-Japan,tie_knot_into_DNA-Quake}, strands of
DNA. Additionally, topoisomerases - proteins which act to alter
the topological state of DNA - are quite common and play a
significant role in cellular processes.

The requirements a knot imposes on a strand are hard to formulate
in a simple way, as ``interactions'' between neighboring strands
can require highly non-local changes in the coil's conformation to
maintain topological state.

That said, the most obvious effect knotting has on a loop is in
the size, commonly measured in terms of radius of gyration,
$R_g^2$. For instance, the loop topologically equivalent to a
circle, called a trivial or $0_1$ knot in professional parlance,
is on average \emph{larger} than the loop of the same length with
any other topology.  In other words, a trivial loop is larger than
the phantom loop, the latter representing topology-blind average
over all loops of a certain length: $\left< R_g^2 \right>_{triv}
> \left< R_g^2 \right>_{phantom} $. This topology-driven
swelling is operational even for very thin polymers, in the
limit when volume exclusion has no effect on polymer coil size. In
this case, the phantom loop's size (which is, once again, average
over all topologies) scales as $N^{1/2}$, while the trivial loop
is larger not merely because of a larger prefactor, but because of
a larger scaling exponent, its size scales as $N^{\nu}$, where
$\nu > 1/2$.  The conjecture, formulated a long time ago
\cite{desCloizeaux_conj}, supported by further scaling arguments
\cite{Quake1,AG_pred}, and consistent with recent simulation data
\cite{Deguchi_2003,swiss_PNAS,Nathan_PNAS}, specifies that the scaling exponent
$\nu$ describing topology-driven swelling of a trivial loop is
exactly the same as the Flory exponent
\cite{Flory_excluded_volume}, which describes swelling driven by
the self-avoidance (or excluded volume): $\nu \approx 0.589
\approx 3/5$.

Equality of scaling exponents for the two cases reflects the
similarity of fractal properties for these systems at very
large $N \gg 1$, because topological constraints result in
self-avoidance of blobs on all length scales above a certain
threshold \cite{AG_pred}.  As we understand much
about self-avoidance \cite{Madras}, and next to nothing
about knots, we would like to exploit the analogy to see if
it yields any insights into knots.  Specifically, it is tempting
to look at the dependence of the unknotted loop size, $\left< R_g^2
\right>_{triv}$, on the number of segments, $N$, not only in the
asymptotic scaling regime of very large $N$, but also the
corrections to scaling at not-so-large $N$.  This is 
particularly important from a practical standpoint, because the
asymptotic scaling limit is barely accessible computationally, and
what one really computes is the value of $\left< R_g^2
\right>_{triv}$ at rather moderate $N$.  Systematic comparison
of $N$-dependencies of $\left< R_g^2 \right>$ for (trivial) knots
and self-avoiding polymers over the wide range of $N$ is the goal
of this paper.

We show that although large $N$ scaling appears to be identical
for trivial knots and excluded volume polymers, their respective approach to
the asymptotic regime is different.  This points obviously to the
limited character of the analogy between the two mechanisms of
swelling, due to volume exclusion and due to topological
constraints.

The plan of the paper is as follows.  We start from a brief
summary of the main results for self-avoiding polymers.
Although these results are widely known, we restate them in the
form most suitable for our purposes.  Next, we present some
heuristic analytical arguments to shed light on why trivial knots
may behave differently then their excluded volume counterparts.  With
this insight in mind, we present our detailed computational
data on the $N$-dependence of $\left<R_g^2 \right>_{triv}$ over
the wide range of $N$.  To obtain data with the necessary
degree of accuracy, it is necessary to make sure that our method
of generating loops is ergodic and unbiased. Although this aspect is of
decisive importance, it is purely technical, and thus it is
relegated to the Appendix.  Up to about section \ref{sec:above} we
mostly review the known results, starting from section
\ref{sec:below}, we present our new findings.

\section{Preliminary Considerations}

\subsection{Swelling driven by self-avoidance: an overview}

To make our work self-contained we now offer a brief review of the
results for the scaling of excluded volume polymers (see further
details in \cite{Madras,AG_Red,Yamakawa}).  We should emphasize
from the beginning that the main properties of the excluded volume
polymer are valid also for loops \cite{Casassa_rings_1965}.  The
simplest model for excluded volume is a system in which $N$ beads,
each of volume $b$, are placed along a loop with mean separation
$\ell$. All other forms of excluded volume, e.g. freely
jointed stiff rods, worm-like filaments, etc., can be mapped to this
simple rod-bead model (see.e.g., \cite{AG_Red}). There are two
scaling regimes, with crossover at the length
\be N^{\ast} \sim \left( \ell^3 / b \right)^2 \ . \label{eq:N*}
\ee
In terms of $N^{\ast}$, the mean squared gyration radius
$\left<R_g^2\right>$ can be written as $\left<R_g^2\right> =
\ell^2 N \rho \left( z \right)$, where the swelling factor $\rho$
depends on the single variable $z=\sqrt{N/N^{\ast}}$.  For 
classical polymer applications, the large $z$ regime is most interesting.
$\rho(z)$ has a branch point
singularity in infinity, its large $z$ asymptotics are dominated by
the factor $z^{2 \nu-1}$; however, if we write $\rho(z) = z^{2
\nu-1} \phi(z)$, then $\phi(z)$ is analytical in infinity and can
be expanded in integer powers of $1/z$.   Accordingly, the large
$N$ asymptotics of $\left<R_g^2\right>$ follow:
\begin{widetext}
\be \left. \left<R^2_g\right> \right|_{N\gg N^{\ast}} \simeq
\ell^2 N^{2 \nu} A \left[1 + k_1
\left(\frac{N^{\ast}}{N}\right)^{1/2} + k_2 \left(
\frac{N^{\ast}}{N}\right)^{1}+ \ldots \right] \ .
\label{eq:self-avoiding-large} \ee
Conversely, in the region $N\ll N^{\ast}$, the approximation for
$\left<R_g^2\right>$ is afforded by the fact that $\rho(z)$ is
analytical at small $z$ and can be expanded in integer powers of
$z$:
\be \left. \left<R^2_g \right> \right|_{1\ll N \ll N^{\ast}}
\simeq \ell^2 N \frac{A^{\prime}}{12} \left[ 1 + k_1^{\prime}
\left(\frac{N}{N^{\ast}}\right)^{1/2} + k_2^{\prime}
\left(\frac{N}{N^{\ast}}\right)^1 + \ldots \right] \ ,
\label{eq:self-avoiding-small} \ee \end{widetext}
where prefactor $A^{\prime}$ should be equal to unity (which
explains why we did not absorb the factor of $1/12$ into
$A^{\prime}$).  Note that the latter result is an intermediate
asymptotics, which means the corresponding region exists only so
long as $N^{\ast} \gg 1$ is large, which means excluded volume is
sufficiently small.

\subsection{Swelling driven by topology: cross-over length}
\label{section:swelling_driven_by_topology}

With this brief summary of results in mind we now set forward,
intending to systematically compare the computational results for
the behavior of trivial knots to the well-understood polymer
with excluded volume.

To look at the analogy between self-avoiding polymers and
trivial knots, it is useful to start, \cite{AG_pred}, by identifying
the cross-over length for knots, an analog of $N^{\ast}$
(\ref{eq:N*}), which we call $N_0$.  For knots, it is
natural to identify the cross-over value of $N$ with the so-called
characteristic length of random knotting, $N_0$; the latter
quantity is known as the characteristic length of the exponential
decay of probability, $w_{triv}(N)$, of formation of a trivial knot
upon random closure of the polymer ends \cite{koniaris_muthu_N0}:
$w_{triv} \simeq \exp(-N/N_0)$.  Depending on the specifics of the
model used,
\cite{koniaris_muthu_N0,Nathan_PNAS,Deguchi_universality_1997},
the critical length varies subtly around $N_0\approx300$.  It is
also clear qualitatively \cite{AG_pred} and seen computationally
\cite{Nathan_PNAS} that this $N_0$ is about the length at which
topological effect on loop swelling crosses over from marginality
at $N<N_0$ to significance at $N>N_0$.  In particular, it is at
$N>N_0$ that the trivial knot begins to swell noticeably beyond
the size of the phantom polymer \cite{Nathan_PNAS}.

\subsection{Swelling driven by topology: above the
cross-over}\label{sec:above}

A number of groups reported observation of the power $\nu \approx
3/5$ in the scaling of trivial
\cite{Deutsch,Deguchi_2003,swiss_PNAS,Nathan_PNAS} and other
topologically simple \cite{Deguchi_2003,swiss_PNAS,Nathan_PNAS}
knots in the region $N>N_0$.

In the works \cite{Deguchi_2003,swiss_PNAS,swiss_macromolecules},
following the idea suggested in \cite{RG_style_fitting}, the $N$
dependence of $\left< R_g^2 \right>_{triv}$ was fitted to the
formula similar to equation (\ref{eq:self-avoiding-large}) for
self-avoiding polymers. No attempt was made at physical
interpretation of the best fit values of the three coefficients
($A$, $k_1$, $k_2$) or the region of $N$ where the fit was
examined. In this sense, fitting with equation
(\ref{eq:self-avoiding-large}) was only used as an instrument to
find the scaling exponent $\nu$, which in these works was found to
be strikingly consistent with the expected value of the
self-avoidance exponent.  A puzzling aspect of the situation is
that, particularly in the work \cite{swiss_PNAS}, the data was fit
to equation (\ref{eq:self-avoiding-large}) not only in the
region $N > N_0$, but across the crossover, starting from about
$N_0/3$ to about $3N_0$ (see also \cite{swiss_macromolecules}).

At present we are aware of no studies which provide a detailed 
comparison of excluded volume and trivial knotting at
modest $N<N_0$. Seeking to further appraise the analogy between
trivial knotting and excluded volume, in the present work we
address the two systems in the region below their respective
crossovers.

\subsection{Swelling driven by topology: below the
cross-over}\label{sec:below}

Formula (\ref{eq:self-avoiding-small}) is the result of
perturbation theory \cite{Yamakawa}, in which conformations with
overlapping segments represent a small part of conformational
space and their exclusion is considered a small correction to
Gaussian statistics. It is tempting to try a similar approach for
knots.  The idea would be to note that at small $N < N_0$, the
probability of a non-trivial knot is small, which implies that
restricting the loop such that it remains a trivial knot excludes only a
small sector of the conformation space which therefore, comprises a
small correction to Gaussian statistics.

Let us try to imagine the realization of this idea.  We want to
find the swelling ratio of the trivial loop: 
\be 
\rho_{0_1} =
\left< R_g^2 \right>_{triv}/\left< R_g^2 \right>_{phantom} \ . 
\ee
We know that the (topology blind) ensemble average over all
knots must, by definition, yield unity for the swelling ratio: 
\be
1 = P_{0_1} \rho_{0_1} + P_{3_1} \rho_{3_1} + P_{4_1} \rho_{4_1} +
\ldots \ , \label{eq:phantom_average_one} 
\ee 
where $P_i$ and
$\rho_i$ are, respectively, the probability and swelling ratio
of the knot $i$.  Our plan is to consider formula
(\ref{eq:phantom_average_one}) as the equation from which we can
determine the quantity of interest, $\rho_{0_1}$: 
\be
\rho_{0_1} = \frac{1 - P_{3_1} \rho_{3_1} - P_{4_1} \rho_{4_1} -
\ldots}{P_{0_1}} \ . 
\label{eq:swelling_ratio_defn}
\ee

\begin{figure}
\centerline{\scalebox{0.4}{
\includegraphics{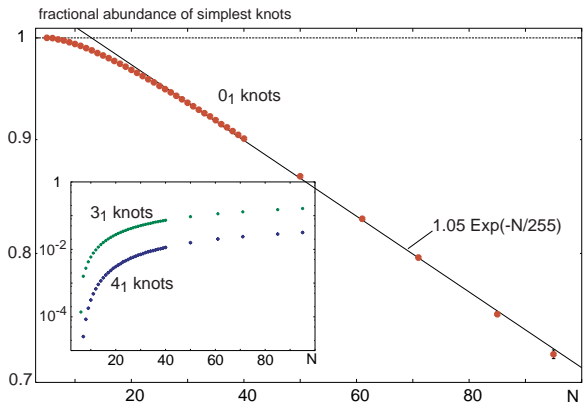}}}
\caption{Fractional abundance of $0_1$, $3_1$, and $4_1$ knots
within the ensemble of all looped polymer chains of fixed
steplength.  The $0_1$ abundance follows the well known,
\cite{koniaris_muthu_N0,probability_DNA_knotting,Deguchi_universality_1997},
exponential decay, $A \exp \left( -N/N_0 \right)$ with decay
length $N_0=255$ and prefactor $A\approx1.05$. Pertinent to the
notion of higher-order knots acting as a perturbation is 
that the abundance of $3_1$ and
$4_1$ knots, seen in the inset, is quite low in the $N\ll N_0$
region. } \label{fig:knot_abundance}
\end{figure}

To this point our consideration is exact, but now we switch to hand
waving arguments and guesses justified by the simulation data. In
the range of small $N$, the ensemble of loops consists mostly of
$0_1$ knots, perturbed slightly by the presence of $3_1$ and
higher-order or more complex knots.  We consider then $N/N_0$ as a small parameter:
$N/N_0 \ll 1$.  Of course, in the case of excluded volume, the similar
limit is better justified, because $N^{\ast}$, equation (\ref{eq:N*}), can
at least in principle, be arbitrarily large, leaving room for the
intermediate asymptotics $1 \ll N \ll N^{\ast}$.  In the case of
knots, $N_0$ is as large as about $300$, but so far we do not know
why it is large, and it seems beyond our control to make it larger.
Accordingly, we cannot speak of an intermediate asymptotics in
a mathematically rigorous way \cite{Barenblatt}.  Nevertheless, we
assume here that the numerically large value of $N_0$ allows 
us hope that the asymptotic argument is possible, and so we 
assume that $N/N_0$ is a small parameter.  We guess then that
higher order knots provide only higher order perturbation
corrections with respect to this parameter, and we neglect their
contributions, simplifying the ensemble by accounting for only
$0_1$ and $3_1$ knots.  In this case, $P_{0_1} + P_{3_1} \simeq
1$.  This is justified by the data presented in Figure
\ref{fig:knot_abundance}, which shows that higher knots are very
rare indeed.  Since we know that $P_{0_1} \simeq \exp (-N/N_0)$,
we can also find $P_{3_1}$.  Given that we consider the $N/N_0 \ll 1$
regime, we must also linearize the exponent, which yields: 
\bea
\rho_{0_1} &\simeq & \frac{1 - (1- P_{0_1}) \rho_{3_1} }{P_{0_1}}
\simeq \frac{1 - \left(1 - e^{-N/N_0} \right)
\rho_{3_1}}{e^{-N/N_0}}  \simeq \nonumber \\
&\simeq& \left( 1 - \left( N/N_0 \right) \rho_{3_1} \right)
\left(1 + N/N_0 \right) \ . \eea

The next step requires thinking about $\rho_{3_1}$.  In principle,
we can come up with a chain of equations, not unlike the BBGKI
chain in the theory of fluids, expressing $\rho_{3_1}$ in terms of
higher knots, etc.   A more practical course is to note that for the lowest
order in perturbation, with respect to the supposedly small
parameter $N/N_0$, since $\rho_{3_1}$ has already the small
($N/N_0$) coefficient in front of it, it is enough to replace
$\rho_{3_1}$ with a constant at $N/N_0 \to 0$.  Thus, to the
lowest order in $N/ N_0 \ll 1$ we get $\left(N /N_0
\right)\rho_{3_1} \simeq \left(N/N_0 \right) c$, where $c$ is a
constant.  We therefore finally obtain \be \rho_{0_1} \simeq 1 +
\left( N/N_0 \right) (1- c) \ , \ee or \be \left< R_g^2
\right>_{triv} \simeq \ell^2 N \frac{1}{12} \left[ 1 + \left(
\frac{N}{N_0} \right) (1- c) \right] \ .
\label{eq:trivial_knot_swelling-perturbation} \ee

The difference between equations (\ref{eq:self-avoiding-small})
and (\ref{eq:trivial_knot_swelling-perturbation}) is immediately
obvious: the former is an expansion in powers of $\sqrt{N}$, the
latter starts from the first power of $N$. The $\sqrt{N}$ term
does not occur in our expansion for knots.  Note that the values
of the $k_i^{\prime}$ coefficients in equation
(\ref{eq:self-avoiding-small}) are known \cite{Yamakawa}, and this
prevents the easy (and incorrect) explanation that $k_1^{\prime}=0$.
As regards the value of coefficient $c$, we do not have at present
an analytical means to calculate it, we will later estimate it
based on the simulation data.  Thus, despite identical scaling
index at large $N$, trivially knotted and excluded volume polymers
exhibit a very different mathematical structure of $N$-dependence in
their respective gyration radii in the region of small $N$.

It is possible that another manifestation of the same difference
is the fact that data in the work \cite{swiss_PNAS} were
successfully fitted to the equation (\ref{eq:self-avoiding-large})
across the crossover region, where this formula for the
self-avoiding polymers is not supposed to work.

Thus, our considerations suggest that there is some fundamental
difference between topology and self-avoidance in terms of their
respective effects on the swelling at moderate $N$.  In what
follows, we present computational tests supporting and further
developing this conclusion.

\section{Model and Simulation Methods}

We model polymer loops as a set of $N+1$ vertices, $\vec{x}_i$,
embedded in $3D$, where $\vec{x}_0=\vec{x}_N$ implies loop
closure. The step between successive vertices, $ \vec{y}_i =
\vec{x}_{i+1}-\vec{x}_{i} $ is constructed either from steps of fixed
length, with probability density
\be 
P(\vec{y}_i)
= \frac{1}{4 \pi \ell^2} 
\delta \left( \left| \vec{y}_{i}\right|-\ell \right) , 
\ee
or Gaussian distributed, with probability density
\be P(\vec{y}_i) = \left( \frac{3}{2 \pi \ell^2}
\right) ^{3/2}\exp \left( - \frac{3 \left|
\vec{y}_{i}\right| ^2}{2 \ell^2} \right) \ . \ee
Note that $\ell$, the ``average'' steplength, is defined,
$\ell^2=\int P(y) y^2 d^3y$. Many methods have been used to
generate loops in computer simulation over the past decade.  A brief
review of the methods is available in Appendix
\ref{appendix:review}, the details of the method implemented 
in this work are presented in Appendix \ref{appendix:ours}.

Once generated, we asses the loop's size by calculating its
radius of gyration
\be R_g^2 = \frac{1}{2N^2} \sum_{i\ne j} \left|
\vec{x}_i-\vec{x}_{j}\right| ^2. \ee
The mean square average radius of gyration seen over all loops is,
$\left< R_g^2 \right>=\frac{1}{12}(N+\beta)l^2$, where $\beta=1$
for fixed steplength loops and $\beta=-1/N$ for loops of gaussian
distributed steplength. Noting that the excluded volume constraint is
maintained by the condition that pair distances be larger than
excluded volume bead diameter, $r_{ij} = \left|
\vec{x}_{i}-\vec{x}_{j}\right| $, $r_{ij} \ge d$, we record the
minimum $r_{ij}$ for each coil, which enables us to ascertain what
maximum diameter of excluded volume, $d$, the loop corresponds to,
\cite{footnote_1}. Finally, we calculate the topological state of
the loop by computing the Alexander determinant, $\Delta(-1)$, and
Vassiliev knot invariants of degree 2 and 3, $v2$ and $v3$, the
implementation of which is described in \cite{Lua_Invariants}. As
the simulation progresses, averages are accumulated in a matrix,
indexed over different knot types and minimum pair distances.  In
the end, we can collect the data to find the gyration radius for
either a particular knot type irrespective of pair distances
(i.e., without volume exclusion), or for a particular excluded
volume value irrespective of topology.

\section{Results}

\begin{figure}
\centerline{\scalebox{0.4}{
\includegraphics{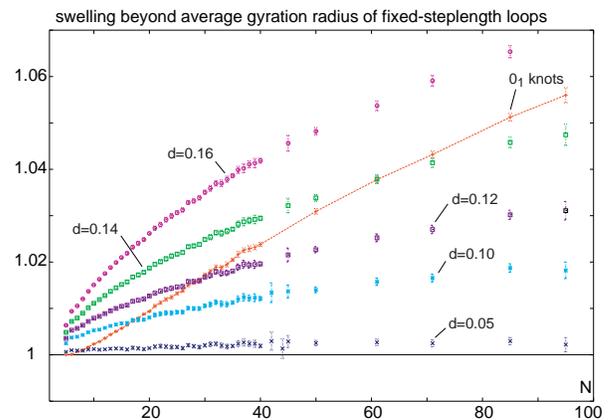}}}
\caption{Direct comparison of excluded volume and trivial knot
swelling, $\rho_{0_1}$, beyond the phantom average size for loops of
fixed steplength.  Excluded volume is formulated in terms of $N$
beads of diameter $d$, each centered at an universal joint between loop
segments.  Exclusion is maintained by prohibiting bead overlap,
$\left| \vec{x}_i - \vec{x}_j\right| \ge d$ for all $i\ne j$.
As discussed in Section \ref{sec:below}, and in contrast to the region 
above their respective crossovers, in the small $N<N_0$ regime, trivial knots follow a functional form
different from that of excluded volume loops. }
\label{fig:compare_exvol_knots}
\end{figure}

\subsection{On the functional form of $N$-dependence of the gyration radius in the moderate $N$ regime}

\begin{figure}
\centerline{\scalebox{0.4}{
\includegraphics{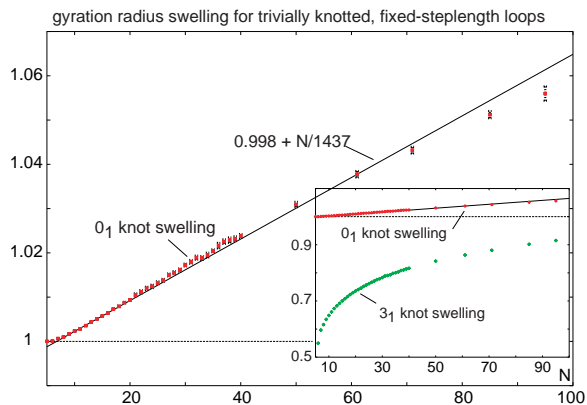}}}
\caption{Average gyration radius data for trivially knotted loops of fixed steplength.
Loops were generated with the conditional probability method described in the appendix.
Swelling of the gyration radius is seen to be linear in the small $N$ regime and can be
understood initially as the result of a perturbation. }
\label{fig:only_knot_swelling}
\end{figure}

Figure \ref{fig:compare_exvol_knots} provides direct comparison
of the computationally determined mean square gyration radius for 
trivial knots and phantom loops with excluded volume 
(averaged over all topologies), in the latter case - for various values of
the bead diameter.  Note that in the figure, the gyration radius is expressed
with the swelling ratio $\rho$, as defined in equation (\ref{eq:swelling_ratio_defn}).
The most striking feature of this figure is the differently shaped
curves of swelling.
The region of intermediate $N$ visible in the figure, $1<N<N_0$, 
shows the plot of trivial knot swelling passing through all 
excluded volume curves.
As seen, the very shape of the $\rho_{0_1}$ curve is different.
Specifically, all curves for the excluded volume loops are bent
downwards, consistent with the presence of the $\sqrt{N}$ terms in
equation (\ref{eq:self-avoiding-small}).  In contrast, the curve
for the topologically restricted trivial loop is very nicely linear.
A fit of the form
\be \rho_{0_1} = 0.998+N/1437 \approx 1 + 0.18 N/N_0 \ , \ee
consistent our estimate, 
equation (\ref{eq:trivial_knot_swelling-perturbation}), 
where $N_0 = 255$,
is shown in Figure \ref{fig:only_knot_swelling}. Note that
deviation from the linear form occurs as $N$ increases.  This is
entirely expected as the crossover to asymptotic swelling of the
gyration radius, $N^{2\nu}/N \sim N^{0.19}$, must occur as $N$
grows beyond $N_0$.

\begin{figure}
\centerline{\scalebox{0.4}{
\includegraphics{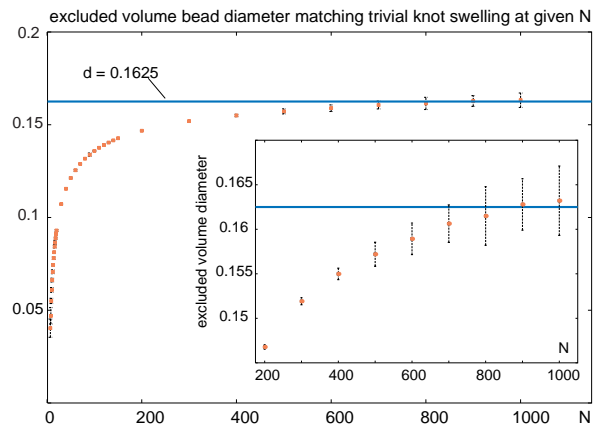}}}
\caption{The excluded volume bead diameter which gives the same $\langle R_g^2\rangle$ swelling as the group of
trivially knotted loops.  Unlike other figures in the publication, loops here are generated
conditionally with gaussian distributed steplength.  This is done for feasibility
reasons, as computationally, gaussian-distributed steps are easier to generate than
loops of fixed steplength.  As seen in the image, the excluded volume diameter seems to saturate at
about $d=0.1625$.  This saturation is consistent with the notion of the trivial knot gyration radius average
approaching the $N^{2\nu}$ asymptotic when $N\gg N_0$.  Although not tested, we expect that
fixed steplength loops would exhibit similar saturation at a specific excluded volume diameter.}
\label{fig:diameter_match}
\end{figure}

\subsection{Which excluded volume diameter matches most closely the topological swelling of trivial knots?}

The cross-over points between curves of trivially knotted loops
and loops with excluded volume in Figure
\ref{fig:compare_exvol_knots} inspired the idea of plotting the
excluded volume diameter at each $N$ whose swelling matches the
swelling of a trivial knot at the same $N$.  As seen in Figure
\ref{fig:diameter_match} this mapping parameter seems to approach
an asymptote at the specific diameter of $d=0.1625$. While at
present it is not computationally feasible to extend the scale of
$N$ to significantly larger values, this asymptotic approach of
trivial knot swelling to loops with excluded volume is consistent
with the similar asymptotic swelling of $N^{2\nu}$ seen in other
work \cite{Deguchi_2003,swiss_PNAS,Nathan_PNAS}.

At the same time, it is interesting to note that although the swelling
parameter due to the excluded volume at $d \approx 0.16$ seems to
fit the topologically driven swelling, the corresponding
characteristic length $N^{\ast}$ (see (\ref{eq:N*})) is
significantly larger than $N_0$.  To see this, we note that
the excluded volume data in figure \ref{fig:compare_exvol_knots} fit
reasonably well to the expression $\rho \approx 1 + 1.71 \sqrt{N}
(d/\ell)^3 = 1 + \sqrt{N/N^{\ast}}$, where, therefore, $N^{\ast} =
0.34 (d/\ell)^6$.  Here, we determined, based on the fit, the
numerical coefficient intentionally left undetermined in formula
(\ref{eq:N*}).  At $d = 0.16 \ell$, we get, therefore, $N^{\ast}
\approx 20000$, which is almost two orders of magnitude greater
than $N_0 \approx 255$.  Alternatively this situation can be seen
by finding the excluded volume diameter for which crossover
length $N^{\ast}$ matches $N_0$: $N^{\ast}=N_0$; the corresponding $d$
equals $d \approx 0.33 \ell$.  It is fairly obvious that this
value of excluded volume does not agree well with the data
presented in figure \ref{fig:diameter_match}.  This discrepancy
possibly points at yet another difference between swelling driven
by topology and excluded volume.

\section{Conclusions}

It seems quite clear from our simulation data that the analogy
between excluded volume and trivial knotting does not hold at loop
sizes smaller than the crossover for knots, $N_0$. The nature of
the swelling function, $\rho (N)$, in this region is yet unknown.
Although our cursory explanation accounts for the trivial knot 
data's linear trend in this regime, the similar parameter
for the size of more complex knots behaves non-linearly, and we
currently have no explanation for this.  A more 
systematic treatment of the problem is badly needed to understand
the size behavior of knots.

That said, our data showing the mapping of excluded volume diameter to trivial knot size seems
to reinforce the notion that asymptotically, the two classes of objects scale with the same power.

We express thanks to R. Lua of the University of Minnesota for the
use of his Knot Analysis routines. We also wish to thank the
Minnesota Supercomputing Institute for the use of their
facilities.  This work was supported in part by the MRSEC Program
of the National Science Foundation under Award Number DMR-0212302.

\appendix

\section{A Brief Review of Loop Generation Methods}
\label{appendix:review}

A number of methods exist and have been used in the literature for
the computational generation of looped polymers.  The goal of
generation methods is to produce statistically representative and
unbiased sets of mutually uncorrelated loops.  The generation of a
random walk is a simple matter.  Steps are chosen with isotropic
probability until the desired length is reached.  Creating random
walks with biased probability, specifically, walks which return to
the origin after a specified number of steps, is a more difficult
task.   As many studies of the topological properties of polymer
chains have been completed, we do not intend to make an exhaustive
summary of all work, but rather in broad strokes summarize the
generation methods used in the field.

All methods used to generate loops can be grouped into two large
categories.  Methods of one group start from some loop
configuration which does not pretend to be random, and then
transform it in some way to randomize the set of steps making the loop.  
Methods of the other group
build more or less random loops from the very beginning.

One of the initial techniques used for the generation of loops is
the dimerization method of Chen, \cite{Chen_1,Chen_3}, in which
smaller sets of walks are joined end to end to form larger walks
or loops. This ``Ring Dimerization'' accepts the joining of
smaller walks with some probability, as self-intersections between
the chains are prohibited.  In addition, if the generated walk is
closed to form a loop, a statistical weight is calculated to
account for loop closure.  Several groups have used this method,
\cite{koniaris_muthu_N0,Deguchi_2002}, usually in the context of
including excluded volume in the topological study.

Other workers, \cite{Deutsch,Deguchi_2003}, start with an initial
loop conformation and then modify it by applying a number of
``elbow'' pivot moves on randomly selected sections of the loop.
Specifically, if the loop is defined by $N$ vertices, $\{
\vec{x}_i \}$, a pivot move is performed by selecting two
vertices, $\vec{x}_j$ and $\vec{x}_k$, and then rotating by a
random angle the intermediate vertices $\vec{x}_{j+1}$ through
$\vec{x}_{k-1}$ about the axis made by $\vec{x}_k-\vec{x}_j$.

A third method in common use, the so-called ``hedgehog'' method
\cite{Vologodskii_hedgehog,swiss_PNAS}, starts by generating $N/2$
pairs of mutually opposite bond vectors.  The resulting set of $N$
vectors has zero sum, and it is tempting reshuffle them
and then use as bond vectors, thus surely obtaining a closed loop.
Unfortunately, such a loop has obviously correlated segments, the
most striking manifestation of which is that the loop has
self-intersections with a large probability of order unity (in
fact, $1/e \approx 0.37$, \cite{combinatorics}; see also a related
scaling argument in \cite{Nathan_PNAS}).  To overcome this, Dykhne
\cite{Vologodskii_hedgehog} suggested imagining all $N$ vectors
plotted from the origin and thus forming something like a
hedgehog, and then randomly choosing pairs of vectors (hedgehog
needles), and rotating the pair by a random angle about their vector
sum. This operation does not change the sum of all $N$ vectors,
which remains zero, and therefore, upon sufficiently many
such operations and upon reshuffling all vectors, one can hope to
obtain a well randomized loop.

The hedgehog method and elbow moves method are in fact quite
similar.  Indeed, in both cases the idea is to rotate some bond
vectors around their vector sum; in the hedgehog method it is done
with pairs of vectors before reshuffling, in the elbow moves
method it is done after reshuffling with a set of subsequent
bonds, but the idea is the same.  In both cases, the evolution of
loop shape can be described by Rouse dynamics, known in polymer
physics (see, e.g., \cite{AG_Red}).  This allows us to make a
simple estimate as to how many moves are necessary in order to
wash away correlations imposed by the initial loop configuration.
Rouse dynamics can be understood as diffusive motion of Fourier
modes.  Since the longest wave Fourier mode has wavelength which
scales as $N$, the longest relaxation time in Rouse dynamics
scales as $N^2$.  This estimate is valid for physical dynamics in
which all segments move at the same time.  Translated into
computational language, this implies that every monomer has to
make about $N^2$ moves, which means that we have to make about
$N^3$ random moves for proper removal of correlations.
 Unfortunately this point is rarely mentioned in the use of these
algorithms, 
(see however, \cite{Deutsch}), and the number of moves between
sampling is generally quite small
, which puts into question the ergodicity of implementations of
this algorithm.

To overcome this problem, we proposed in \cite{Nathan_PNAS}
another method which we call the method of triangles, which does
not involve any relaxation.   In this method, we generate $N/3$
randomly oriented triplets of vectors with zero sum, reshuffle 
them, and connect them
head-to-tail, thus obtaining a loop.  As we shall explain in
another publication, this method produces loops with insignificant
correlations when $N$ is larger than a hundred or so.

Since our major attention in this article is the range of
relatively small $N$, we have to resort to a computationally more
intensive, but reliably unbiased method based on conditional
probabilities.  The idea is to generate step number $i$ in the
loop of $N$ steps using the conditional probability that the given
step arrives to a certain point provided that after $N-i$ more
steps the walk will arrive at the origin.  This method was
suggested and implemented for Gaussian chains in \cite{volog}.
Here, we apply it for the loops with fixed step length.

\begin{widetext}

\section{Generation of loops with fixed steplength using the conditional probability method}\label{appendix:ours}

\subsection{Derivation of the Conditional Probability
Method}

A walk is composed of $N$ steps between $N+1$ nodes, a step from
nodes $\vec{x}_i$ to $\vec{x}_{i+1}$ having normalized
probability, $g(\vec{x}_i,\vec{x}_{i+1},1)$.  The probability for
a random walk composed of $N$ such steps is described by the Green
function which ties the steps together,
\be G(\vec{x}_0,\vec{x}_N,N)
=\int g(\vec{x}_1-\vec{x}_0)g(\vec{x}_2-\vec{x}_1)...g(\vec{x}_N-\vec{x}_{N-1})
d\vec{x}_1d\vec{x}_2 ... d\vec{x}_{N-1}
\label{eq:multiple_steps}\ee
Note that in this notation the walk stretches from $\vec{x}_0$ to
$\vec{x}_N$. The specifics of integration depend on the sort
of steps which are being taken.  At times, these integrations can
be difficult to evaluate.  In such cases the convolution theorem
can be of some utility.  Suppose that the Fourier transform and
inverse is defined in the usual way,
\be
\begin{array}{cc}
g_{\vec k} =\beta \int  g(\vec x) \exp \left[ i \vec{k} \cdot \vec{x} \right] d\vec{x} \\
g(\vec x) = \beta \int g_{\vec k} \exp\left[ -i \vec{k} \cdot \vec{x}\right] d\vec{k} .
\end{array}
\label{eq:fixed_step_intergration-specific}
\ee

Note that in this formulation $\beta=(2\pi)^{-3/2}$.  The
convolution theorem allows for the following expression for $N \ge
2$,
\be
G(\vec{x}_0,\vec{x}_N,N) =
        (1/\beta)^{N-2} \int  (g_{\vec k})^N \exp\left[ - i \vec{k} \cdot (\vec{x}_N - \vec{x}_0)\right] d\vec{k} \ .
\label{eq:step_convolution} \ee

If steplength is fixed to a certain distance, $\ell$, the
probability distribution and its fourier transform are expressed,
\be
\begin{array}{cc}
        g(\vec{x}_0,\vec{x}_1,1)_{fixed}= \frac {\delta(\left| \vec{x}_1-\vec{x}_0\right| -\ell)}{4 \pi l^2}\\
        \\
        g_{\vec{k}} = \beta \frac{\sin\left(k \ell \right)}{k \ell} \ ,
\label{eq:FT_fixed_step}
\end{array}
\ee Using equations (\ref{eq:step_convolution}) and
(\ref{eq:FT_fixed_step}), along with differential volume
$d\vec{k}=2 \pi k^2 dk d(\cos \theta)$, the probability
distribution for a walk of $N$ fixed-length steps spanning the
displacement $\vec{x}_N-\vec{x}_0$ is, \be
G(\vec{x}_0,\vec{x}_N,N)_{fixed} =
        \beta^2
        4 \pi \int^{\infty}_0 {\left(\frac{\sin\left[k \ell \right]}{k \ell} \right)}^N
        \frac{\sin\left[ k \left| \vec{x}_N - \vec{x}_0\right| \right]}{k \left|
        \vec{x}_N - \vec{x}_0\right| }  k^2 dk \ .
\ee

If we use the definition of $\beta$ and express Sine terms as
exponentials, also using $d=\left| \vec{x}_N - \vec{x}_0\right|
/\ell$ then,
\be
G(\vec{x}_0,\vec{x}_N,N)_{fixed} =
        \frac{1}{2 \pi^2}
        \int^{\infty}_0 \frac{\left(\exp\left[i k \ell\right]- \exp\left[-i k \ell\right]\right)^N
        \left( \exp\left[i k \ell d\right]-\exp\left[-i k \ell d\right] \right)}{(2ik \ell)^{N+1}d}k^2dk  \ .
\ee
Then using the Newton binomial $(x+y)^N=\sum_{m=0}^N {N \choose m}
x^{N-m}y^m$, where, ${N \choose m}=\frac{n!}{(n-m)!m!},$ yields a
shiny prize, an analytically tractable expression:
\be
G(\vec{x}_0,\vec{x}_N,N)_{fixed} =
        \frac{1}{\pi^2} \frac{1}{2^{N+2}i^{N+1}\ell^{N+1}d}
        \int^{\infty}_{0} 
	\sum_{m=0}^N {N \choose m} 
	\frac
	   {(\exp\left[i k \ell \right])^{N-m}(-\exp\left[-i k \ell\right])^m
        	(\exp\left[i k \ell d\right]-\exp\left[-i k \ell d\right])}
	   {k^{N-1}}
	dk \ .
\ee
%
%
At this point two further simplifications are made.  The first is
to extend the integration from $-\infty$ to $\infty$, as the
integrand is even on the real axis (with proper incorporation of
the factor of $1/2$). The second simplification is to integrate
over the dimensionless number, $\kappa=k \ell$.  Note that the
dimension of the integral remains $1/volume$.
\be
G(\vec{x}_0,\vec{x}_N,N)_{fixed} =
        \frac{1}{\pi^2} \frac{N!}{2^{N+3}i^{N+1}\ell^3d}
        \int^{\infty}_{-\infty} \sum_{m=0}^N
         \frac{{(-1)}^m}{(N-m)!m!}
        \frac{\exp\left[i \kappa( N-2 m+d)\right]-\exp\left[i
        \kappa (N-2m-d)\right]}{\kappa^{N-1}}d\kappa \ .
\ee
The integral which remains can be evaluated as a contour integral in the
complex plane.  The contour along the real axis is chosen with a small bump
in the $+i$ direction at $\kappa=0$.  The upper or lower arch is chosen
according to Jordan's Lemma.  The residue at $\kappa=0$ is obtained by
Taylor expanding the exponent to resolve the coefficient corresponding to
the $\kappa^{-1}$ term, which is the definition of a residue.  The result follows,
\be
\int_{-\infty}^{\infty}\frac{\exp\left[i \alpha \kappa\right]}{\kappa^{N-1}}d\kappa =
        \left\{
                \begin{array}{ccc}
                        0 & \rm{if} & \alpha \ge 0\\
                        -2 \pi i \left( \frac{1}{(N-2)!}(i \alpha)^{N-2}  \right) & \rm{if} & \alpha < 0
                \end{array}
        \right. \ .
\ee 
Integration winnows the sum considerably,
the final result is,
\be
G(\vec{x}_0,\vec{x}_N,N)_{fixed} =
        \frac{N(N-1)}{2^{N+2}\pi l^3 d }
        \left(J_1(N,d)-J_2(N,d)\right) \ , 
\label{eq:fixed_prob_w_J}
\ee
where \be
J_1(N,d)=\sum^N_{m>(N+d)/2}\frac{(-1)^m}{(N-m)!m!}(N-2m+d)^{N-2} \
, \ee and \be
J_2(N,d)=\sum^N_{m>(N-d)/2}\frac{(-1)^m}{(N-m)!m!}(N-2m-d)^{N-2} \
. \label{eq:sum_ennumeration} \ee
A table of probabilities can then be composed.  Note however that
the probability is defined on intervals over $d$, listed in the
right column below. \be
\begin{array}{cc}
G(\vec{x}_0,0,3)_{fixed} = \left\{ \begin{array}{cc}
                \frac{1}{8\pi \ell^3 d} & d \in \left[0,2\right]
        \end{array} \right. \\
\\
G(\vec{x}_0,0,3)_{fixed} = \left\{ \begin{array}{cc}
                (1)/(8 \pi \ell^3) & d \in \left[0,1\right] \\
                (3-d)/(16 \pi d \ell^3) & d \in \left[1,3\right]
        \end{array} \right. \\
\\
G(\vec{x}_0,0,4)_{fixed} =  \left\{ \begin{array}{cc}
                (8-3d)/(64\pi \ell^3) & d \in \left[0,2\right] \\
                (d-4)^2/(64\pi \ell^3 d) & d \in \left[2,4\right]
        \end{array} \right. \\
\\
G(\vec{x}_0,0,5)_{fixed} = \left\{ \begin{array}{cc}
                 (5-d^2)/(64 \pi \ell^3)& d \in \left[0,1\right] \\
                 (2d^3-15d^2+30d-5)/(192 \pi \ell^3 d)& d \in \left[1,3\right] \\
                -(d-5)^3/(384 \pi \ell^3 d) & d \in \left[3,5\right]
        \end{array} \right. \\
\\
G(\vec{x}_0,0,6)_{fixed} = \left\{ \begin{array}{cc}
                (5d^3-24d^2+96)/(1536 \pi \ell^3)& d \in \left[0,2\right] \\
                (-5d^4+72d^3-360d^2+672d-240)/(3072 \pi \ell^3 d) & d \in \left[2,4\right] \\
                (d-6)^4/(3072 \pi \ell^3 d)) & d \in \left[4,6\right]
        \end{array} \right. \ .
\\
\end{array}
\ee

These piecewise-defined probability distributions approach the shape of the corresponding
quantity for gaussian distributed steplength,
\be
        G(\vec{x}_0,\vec{x}_N,N)_{gaussian}=
                \left(\frac{3}{2\pi N \ell^2}\right)^{3/2}
                \exp\left[-\frac{3}{2N  \ell^2}(\vec{x}_N-\vec{x}_0)^2\right].
\label{eq:gaussian_walk} 
\ee 
Due to the complexity and computational expense of the conditional method, 
and noting the apparent similarity of the two curves, one might be tempted to substitute 
the Gaussian formulation, equation (\ref{eq:gaussian_walk}), when $N$ is above some 
threshold, $N>N_c$.  Our own 
experience with this approximation leads us to discourage the intermingling of the two 
distributions. When included, at even the large $N_c=30$, a sharp discontinuity in the 
curve of curve for $\rho_{0_1}$ vs $N$ (Figure (\ref{fig:only_knot_swelling})) 
was visible at $N_c$.
We hypothesize that substitution of the Gaussian formulation, equation (\ref{eq:gaussian_walk}), 
for the fixed-step formulation, equation (\ref{eq:fixed_prob_w_J}), allows for slightly more
inflated loop conformations and thus leads to a discontinuity when the approximation 
is used in the simulation code at $N>N_c$.  

\end{widetext}

\subsection{Implementation of Conditional Probability Method}

Generation of a random walk which is looped, i.e. $\vec{x}_N -
\vec{x}_0=0,$ can be achieved with the use of the already derived
equations.  Imagine that a walk of $N+M$ steps is underway and $M$
steps have already been taken.  This means that a walk of $N$
steps remains, which starts at the present location, $\vec{x}_0$,
and finishes at the starting point, $\vec{x}_N$.  The probability
distribution for the next step, from $\vec{x}_0$ to $\vec{x}_1$,
can then be written,
\be
P(\vec{x}_0|\vec{x}_1) =
    \frac
        {G(\vec{x}_0,\vec{x}_1,1)G(\vec{x}_1,\vec{x}_N,N-1)}
        {G(\vec{x}_0,\vec{x}_N,N)}
        \label{eq:conditional_probability}
\ee
In principle one could generate new steps with probability
isotropic in direction, accepting them with conditional
probability defined by equations (\ref{eq:conditional_probability})
and (\ref{eq:fixed_prob_w_J}) or (\ref{eq:gaussian_walk}).  In
the interest of efficiency, a better method is to generate random
steps within these probability distributions. Now discussed is the
way to transform a flat random distribution (that produced by the
UNIX math function drand48() for example) into the distribution above.
If the flatly distributed variable is $q$, ie $P(q)=1$ on
$[0,1)$, $0$ elsewhere, the following equation, with
$\vec{d}=\vec{x}_N-\vec{x}_1$,  defines the transform to the
conditional distribution above, $G(\vec{x})$,
\be \int_{0}^{q}P(q^{\prime})d(q^{\prime}) =
        \int_{0}^{f(q)} \frac
                {G(\vec{x}_0,\vec{x}_1,1)G(\vec{d},0,N-1)}
                {G(\vec{x}_0,\vec{x}_N,N)} d(\vec{d}),
\label{eq:transform_drand}
\ee
In this statement of normalization, the function of importance
is $f(q)$, which defines the way the two probability distributions are made equal.

In principle the problem is now solved.  A complete set of
probability distributions for walks of fixed or gaussian
steplength has been defined, and the the formula which maps that
distribution to a flat, machine-generated distribution has also
been expressed.  If the form of equation
(\ref{eq:transform_drand}) is simple enough, meaning relatively
small $N$, the integral equation can be solved directly for
$f(q)$.  In practice however, $N>5$ is an interesting regime and a
different technique must be used to obtain $f(q)$.

\begin{figure}
\centerline{\scalebox{0.4}{
\includegraphics{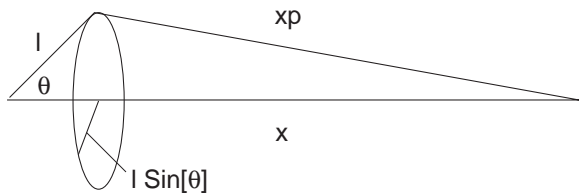}}}
\caption{This geometry is used in the implementation of the 
conditional probability loop generation method.}
\label{fig:diagram}
\end{figure}

For the case of finishing a random walk of fixed length steps,
$\ell$, which is $\vec{x}$ away from the ending point, and has $N$
steps alloted to get to that point, we use the geometry shown in
figure (\ref{fig:diagram}).  In this diagram $\vec{x}_p$ is the
new distance away from the endpoint after the present step is
taken. Thus the expression above becomes, 
\be
\int_{0}^{q}P(q^{\prime})dq^{\prime} = \int_{0}^{f(q)}
        \frac{G(\vec{l},1) G(\vec{x_p},N-1)}{G(\vec{x},N)} d(\vec{x_p}) \ ,
        \label{eq:specific_transform}
\ee
where, for convenience, the following syntax is used, $G(\vec{b},0,N) = G(\vec{b},N)$.

Of course the single step $G(\vec{d},1)$ is a delta-function,
$\delta ( | \vec{d} | -\ell) / 4 \pi \ell^2$, so the
integration over $d(\vec{x_p})$ occurs over most or all of the
spherical shell created by the possible orientations of $\ell$. Integration
over the shell (about the axis made by $\vec{x}$) is performed in
``rings,'' each ring having circumference $2\pi \ell \sin[\theta]$,
and width, $\ell d(\theta)$, with resulting differential area, $dA = 2 \pi
\ell^2 \sin[\theta] d(\theta)$.  $\theta$ is integrated over the
range, $[0,\pi]$.

It should be apparent that, 
${x_p}^2 = x^2+\ell^2+2x\ell\cos[\theta]$. 
This yields the differential transform, 
$\sin[\theta]d(\theta) = (x_p / x \ell)d(x_p)$.  
This simplification allows the integration of equation
(\ref{eq:specific_transform}) in the following way,
\be \int_{0}^{q}P(q^{\prime})dq^{\prime} = \frac{1}{2 \ell x
G(x,N)} \int_{min}^{f(q)}
        G(\vec{x_p},N-1) x_p d(x_p),
\label{eq:working_transform}
\ee
This expression is normalized to $1$ if integrated over
appropriate $x_p$  bounds. In most cases, those bounds are
$[x-\ell,x+\ell]$, although the physical limit on the upper bound,
$x_p \le(N-1)\ell$ is necessary to keep the walk from straying too
far from the origin. Additionally, if the walk is very close to
the origin, $x<\ell$, the integration bounds, $[\ell+x,\ell-x]$,
are used.

As Equation (\ref{eq:sum_ennumeration}) for fixed steplength
probability is defined as a polynomial, integration of that
polynomial, described by Equation (\ref{eq:working_transform}), can
be performed exactly within simulation computer code, and the 
resulting equation for $f(q)$ solved numerically.  In practice
we use the Gnu Multiple Precision library to represent the
polynomial coefficients and values as {\it rational} numbers. 
From a computational standpoint this is significantly more 
expensive than representing coefficients
as double floating point, but using rationals allows us to represent all outputs
of the polynomial with great accuracy, the goal of this simulation
method.  At a relatively small number of steps the coefficients become 
quite small, for example at $N=15$, in the
region $x\in[13,15]$, equation (\ref{eq:sum_ennumeration}) reads,
$\frac{-(d-15)^{13}}{40809403514880( \ell^3 d \pi)}$.  
We feel the 
need in this routine to retain accuracy when performing operations 
such as $P-Q$, where $P\gg1$ and $Q\gg1$ but $(P-Q)\ll P,Q$. 
In order to retain
the accuracy of the conditional formulation it was imperative to
perform this rational number algebra.  For the interested reader
we provide a table of these polynomial coefficients as supplementary 
materials.


%


\begin{thebibliography}{99}
\bibitem{footnote_1}
Use of this computational trick means that ensembles for similar excluded
volume diameters suffer from significant correlation.  Nevertheless, given
the qualitative nature of our investigation of excluded volume we feel this
is not a dangerous correlation to include.
\end{thebibliography}
\end{document}


\title{On the Limits of Analogy Between Self-Avoidance and Topology-Driven Swelling of Polymer Loops}

\author{N.T. Moore, A.Y. Grosberg}
\affiliation{Department of Physics, University of Minnesota, Minneapolis, MN 55455, USA}
\date{\today}

\maketitle

\section{Supplementary Materials}

\subsection{Polynomial Description}

In the associated work, we derive expressions for the conditional probability of a random walk, 
$G(d \ell,n)_{fixed}$, with end to end distance $d \ell$ from $n$ steps of fixed length 
$\ell$. In these supplementary materials, we present these 
probabilities in polynomial form, making them computationally more accessible.

First, note that the sum indices in Equations B11 and B12 lead to a piece-wise definition of 
the functions (this is seen in Equation B13).  The function, $G(d \ell,n)_{fixed}$,  
is defined on $(n+1)/2$ intervals if $n$ is odd, and $n/2$ intervals if $n$ is
even. All intervals span distance in $d$ of $2$, with exception of the first interval 
for odd $n$, $d \in [0,1]$.

Use of a computer algebra system allows for the expansion of equation B10 into the form, 
\begin{equation}
\frac{1}{d \pi l^3} \sum_{i=0}^{n-2} a_{i} d^{i},
\end{equation}
where the set of $a_i$ coefficients are rational numbers. Note however that there will be roughly
$n/2$ such sums for each $n$ as described in the previous paragraph.  It is therefore useful to 
think of the sets of coefficients as list elements, $a_{n,j,i}$, where $n$ refers to the number 
of steps in the walk, $j$ refers to the interval in $d$ of definition, and $i$ to the specific 
power of $d$ in the polynomial.

Polynomials are defined in the attached {\it fractions32.txt} and {\it fractions101.txt} coefficient 
files on the interval of $n \in [2,n_c]$, in this case $n_c=32$ and $n_c=101$ respectively. 
The first element in each file is $n_c$.

Subsequent file elements are the $a_{n,j,i}$ coefficients, delimited by commas, and provided 
as rational numbers.  Coefficients are listed by a nested iteration, first by step number $n$, 
then by interval $j$, finally by polynomial index $i$. Were $n_c=3$, the file would read:

\begin{equation}
\underbrace{ 3 }_{n_c}, 
\underbrace{ 1/8 }_{a_{n=2,j=0,i=0}}, 
\underbrace{ 0}_{a_{n=3,j=0,i=0}}, 
\underbrace{1/8}_{a_{n=3,j=0,i=1}}, 
\underbrace{ 3/16,-1/16 }_{a_{n=3,j=1,i}} 
\end{equation}